\input{aipcheck}
\documentclass[final]{aipproc}
\layoutstyle{6x9}
\usepackage{amsmath}
\begin{document}
\title{Spin, Orbital, and Spin-Orbital Polarons in Transition Metal Oxides}
\author{Krzysztof Wohlfeld}{address={Marian Smoluchowski Institute of Physics, Jagellonian University, \\Reymonta 4, PL-30059 Krak\'ow, Poland, \\ and Max-Planck-Institut f\"ur Festk\"orperforschung, \\Heisenbergstrasse 1, D-70569 Stuttgart, Germany}}
\begin{abstract}
I give a brief overview of a polaron formation in three distinct 
transition metal oxides:
(i) spin polaron when a hole is added to the antiferromagnetic (AF) ordered plane in La$_2$CuO$_4$, 
(ii) orbital polaron when a hole is added to the alternating orbital (AO) ordered plane in LaMnO$_3$, 
and (iii) spin-orbital polaron when a hole is added to the AF and AO ordered plane in LaVO$_3$. 
Comparison of the distinct features of the above polarons can shed some light on the basic
differences between the experimental phase diagrams of the lightly doped transition metal oxides
La$_{2-x}$Sr$_x$CuO$_4$, La$_{1-x}$Sr$_x$MnO$_3$, and La$_{1-x}$Sr$_{x}$VO$_3$. 
\end{abstract}
\maketitle
\section{INTRODUCTION}
The doped transition metal oxides have very rich phase diagrams which
are fingerprints of the spectacular physics present in these strongly correlated
electron systems \cite{Ima98}. In particular: (i) La$_{2-x}$Sr$_x$CuO$_4$ has
an antiferromagnetic (AF) and Mott instulating ground state only for very low doping $x \in (0,0.02]$
although it is a high-temperature superconductor with optimal doping $x\sim 0.15$ \cite{Ima98}, 
(ii) La$_{1-x}$Sr$_x$MnO$_3$ has a plane with a ferromagnetic (FM) $e_g$ alternating orbital 
(AO) and Mott insulating ground state in the lightly doped regime $x \in (0, 0.18]$ \cite{End99}, 
and (iii) La$_{1-x}$Sr$_{x}$VO$_3$ has a plane with an AF and a $t_{2g}$ AO Mott insulating ground state in 
the lightly doped regime $x \in (0, 0.178]$ \cite{Fuj05}. In the present paper we
try to shed  some light on the distinct features of the phase diagrams of 
these three lightly doped transition metal oxides.

From the theoretical point of view the description of the lightly doped transition 
metal oxides is relatively easy in the extreme case of only one hole doped into the 
half-filled state \cite{Mar91} and in what follows we reduce our studies to this
limit only. Then the motion of such a single hole added to the half-filled 
Mott insulating ordered ground state is strongly renormalized as it couples to 
the excitations of the ordered state, magnons in the AF state and orbitons
in the AO state \cite{Mar91, Bri00}. This means that a {\it polaron} is formed: 
(i) in La$_2$CuO$_4$ with one hole in the AF ground state -- a {\it spin} polaron \cite{Mar91}, 
(ii) in LaMnO$_3$ with one hole in the AO ground state -- an {\it orbital} polaron \cite{Bri00}, and
(iii) in LaVO$_3$ with one hole in the AO and AF ground state -- a {\it spin-orbital} polaron \cite{Woh09}. 
In the next three chapters, using the self-consistent Born approximation (SCBA) \cite{Mar91} applied
to the polaron formulation of the respective $t$-$J$ model, we compare features of these three different
polarons \cite{Ber09}.

\section{SPIN POLARON}
\begin{figure}[t]
\includegraphics[width=0.46\textwidth]{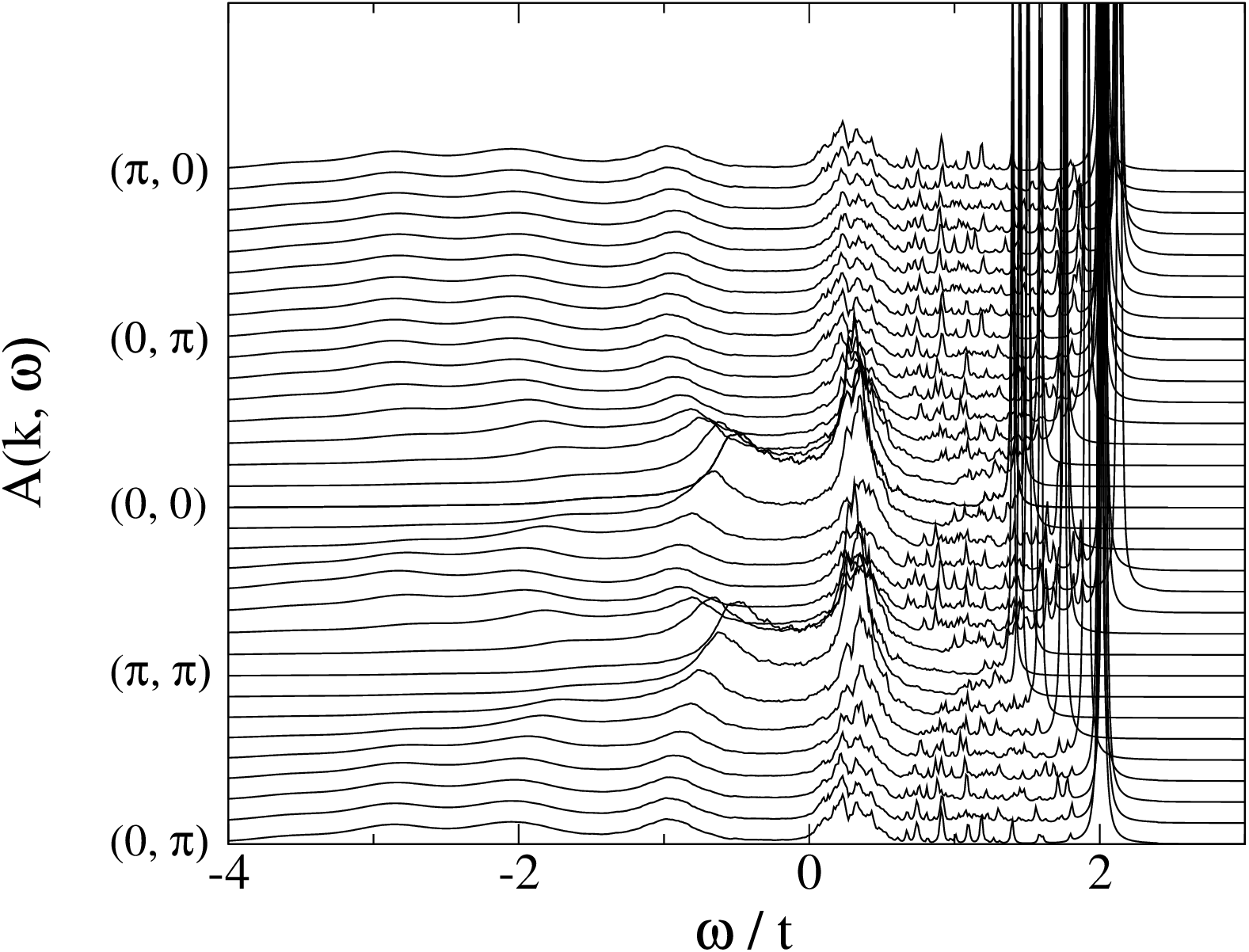}
\caption{\small{
Spectral density $A({\bf k}, \omega)$ of the model Eq. (\ref{eq:1}) with $J=0.4t$ 
along the particular directions of the 2D Brillouin zone.
} }
\label{fig:1}
\end{figure} 

It is believed that the basic effective model which describes the important physics present 
both in the undoped La$_2$CuO$_4$ and in the lightly doped La$_{2-x}$Sr$_x$CuO$_4$ 
is the two-dimensional (2D) $t$-$J$ model \cite{Cha78},
\begin{equation}
\label{eq:1}
H_S=-t \sum_{\langle i,j \rangle, \sigma}   \left(\tilde{c}^\dag_{i\sigma} \tilde{c}_{j\sigma}
+ {\rm H.c.}\right)+ J \sum_{\langle i, j \rangle} {\bf S}_i \cdot {\bf S}_j, 
\end{equation}
where ${\bf S}_i$ are spin $S=1/2$ operators and
the constrained operators $\tilde{c}^\dag_{i\sigma}=c^\dag_{i\sigma}(1-n_{i\bar{\sigma}})$
allow for the hopping only in the restricted Hilbert space with no double occupancies.
The superexchange energy scale is $J=4t^2/U$ where $U$ is the {\it effective} repulsion between 
two electrons with opposite spins on the same Cu site and $t$ is the effective hopping between the Cu ions. 

In the undoped case the ground state of the model is the 2D AF ordered state -- this can
be easily seen by noting that the kinetic term in Eq. (\ref{eq:1}) does not contribute
in the half-filled case and the $t$-$J$ model reduces then to the Heisenberg model.
On the other hand, in the case of one hole doped into the AF ground state the model
Eq. (\ref{eq:1}) can be reduced to the polaron-type model with the quadratic terms
representing magnon spectrum and the polaron-type interaction between the holes
and the magnons \cite{Mar91}. Such a model can be easily solved using the SCBA method 
\cite{Mar91} and the hole spectral functions can be calculated from the Green's functions.

We solved the SCBA equations numerically on a mesh of $16 \times 16$ points
-- the results for the realistic case of $J=0.4t$ are shown in Fig. \ref{fig:1}.
We see a well-developed dispersive quasiparticle peak on the right hand side of the spectrum which
suggests that the polaron is formed. Since microscopically the polaron is formed
due to the coupling between the hole and magnons we call it a {\it spin} polaron.
Besides, we note that the excited states are almost entirely different than those
found in the classical Ising case (the so-called ladder spectrum \cite{Mar91}).
This is particularly pronounced for some values of momentum ${\bf k}$ such as e.g.
${\bf k} = (0,0)$ or ${\bf k} = (\pi,\pi)$.

\section{ORBITAL POLARON}
\begin{figure}[t]
\includegraphics[width=0.46\textwidth]{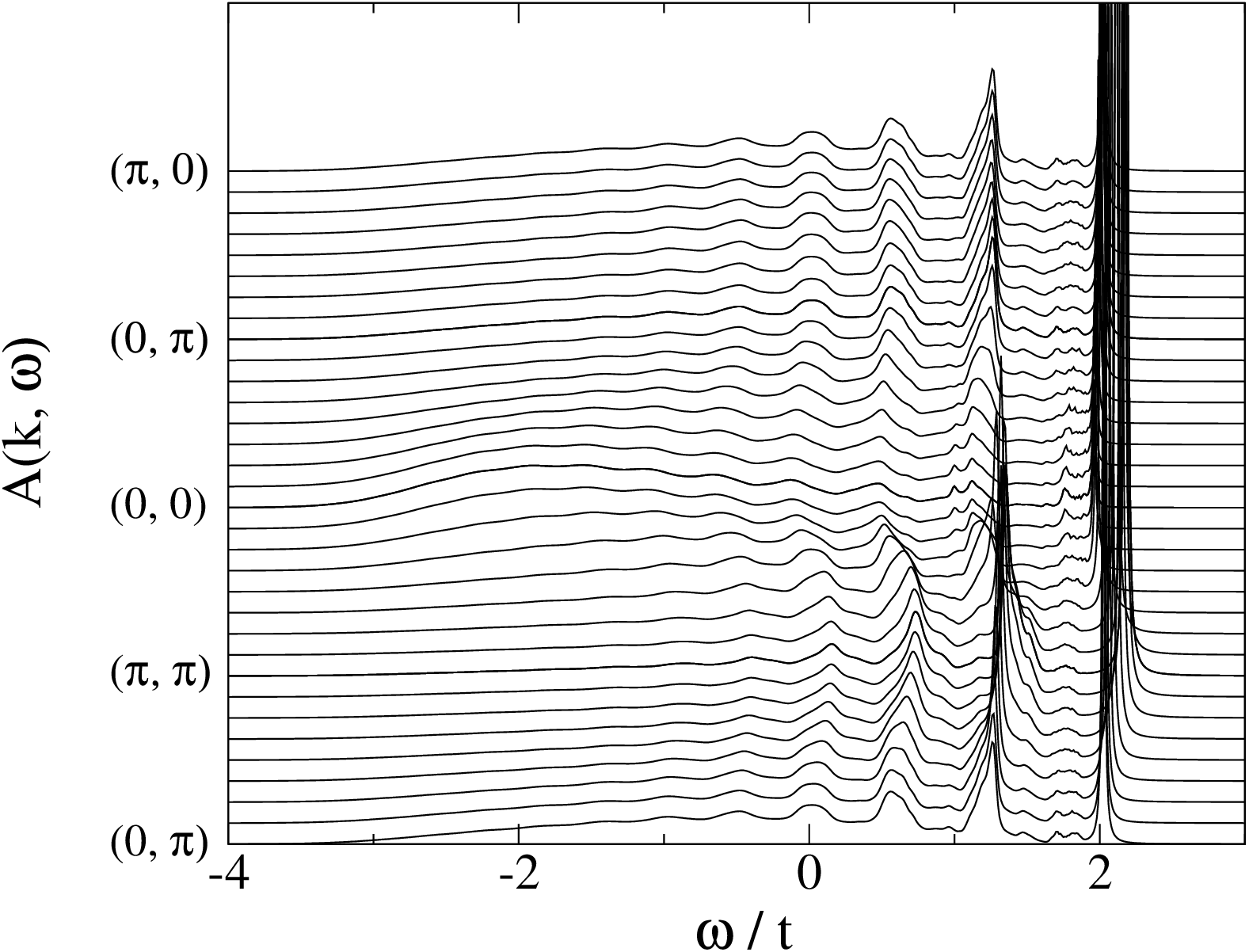}
\caption{\small{
Spectral density $A({\bf k}, \omega)$ of the model Eq. (\ref{eq:2}) with $J=0.4t$ 
along the particular directions of the 2D Brillouin zone.
} }
\label{fig:2}
\end{figure} 
A different situation occurs in the undoped LaMnO$_3$ and its doped counterpart La$_{1-x}$Sr$_x$MnO$_3$:
here the partially filled $e_g$ orbitals are degenerate and the orbital degrees should be taken
into account. However, in the lightly doped case the 2D ferromagnetic state is stable 
and consequently the spin degrees of freedom can be integrated out. Thus, one arrives 
at the following ${\it orbital}$ $t$-$J$ model for the $(a,b)$ plane \cite{Fei05, Bri00}, 
\begin{align}
\label{eq:2}
H_O=&-\frac{1}{4}t \sum_{\langle i,j \rangle}   \left[3\tilde{x}^\dag_{i} \tilde{x}_{j}
+\tilde{z}^\dag_{i} \tilde{z}_{j} \mp \sqrt{3}\left(\tilde{x}^\dag_{i} \tilde{z}_{j}
+\tilde{z}^\dag_{i} \tilde{x}_{j}\right) + \mbox{H.c.}\right] 
\nonumber \\
&+\frac{1}{8}J \sum_{\langle i, j \rangle} \left[3T^x_i T^x_j+T^z_i T^z_j \mp \sqrt{3}\left(T^x_i T^z_j
+T^z_i T^x_j\right)\right], 
\end{align}
where $x^\dag_{i}|0\rangle = \frac{1}{\sqrt{2}} |x^2-y^2\rangle_i$, 
$z^\dag_{i}|0\rangle = \frac{1}{\sqrt{6}} |3z^2-r^2\rangle_i$, 
the $-$($+$) signs denote the bonds along the $a$ ($b$) direction,
and tilde
denotes the hopping in the Hilbert space with no double occupancies. 
Besides, ${\bf T}_i$ are pseudospin 
$T=1/2$ operators with $T^z_i=(\tilde{n}_{ix}-
\tilde{n}_{iz})/2$, the superexchange energy scale is $J=4t^2/U$ where $U$ is the effective repulsion
between electrons in the ${}^6A_1 $ state, and $t$ is the effective hopping between
the Mn ions. 

This time, in the undoped case the ground state of the model is the 2D AO ordered state 
formed by the $(|x\rangle+|z\rangle)/\sqrt{2}$ and $(|x\rangle-|z\rangle)/\sqrt{2}$ orbitals.
When one hole is doped into such an AO ground state the model Eq. (\ref{eq:2}) can be again 
reduced to the polaron-type model with the quadratic terms
representing {\it orbiton} spectrum and the polaron-type interaction between the holes
and the orbitons \cite{Bri00}.

We solved the respective SCBA equations numerically on a mesh of $16 \times 16$ points
-- the spectral function for the realistic case of $J=0.4t$ \cite{Bri00} is shown in Fig. \ref{fig:2}.
As in the spin case there is a well-developed quasiparticle peak on the right hand side 
of the spectrum which suggests that the polaron is formed. However, it is an {\it orbital}
polaron since it describes a hole dressed by orbiton excitations. Moreover, 
the quasiparticle peak has almost no dispersion and the excited states 
resemble the ladder spectrum \cite{Mar91} which suggests that 
the orbital polarons are much more "classical" than the spin ones.

\section{SPIN-ORBITAL POLARON}
In the undoped LaVO$_3$ and in the lightly doped La$_{1-x}$Sr$_x$VO$_3$ the situation
is much more complex than in the cuprates or manganites: here both the spin and orbital
degrees of freedom should be taken into account as the spin degrees of freedom
form the AF order in the undoped plane and cannot be integrated out as in
the manganites \cite{Kha01}. Thus, one needs to consider the full spin-orbital $t$-$J$ model 
with $t_{2g}$ orbital degrees of freedom \cite{Kha01, Woh08}. Furthermore, 
in the case of the $t$-$J$ models with $t_{2g}$ orbital degrees of freedom
we have to supplement such models with the frequently neglected three-site terms \cite{Dag08}.
Thus we arrive at the following strong-coupling model for the $(a,b)$ planes of
the cubic vanadates \cite{Woh09},
\begin{equation}
\label{eq:3}
H_{SO}=H_t+H^{(1)}_J+H^{(2)}_J+H^{(3)}_J+H_{3s}.
\end{equation}

The first term in the above equation is the kinetic term \cite{Woh09},
\begin{align}\label{eq:ht}
H_t &= -t \sum_{i,\sigma} P \left( \tilde{b}^\dag_{i\sigma} \tilde{b}_{i+\hat{a}\sigma}
+\tilde{a}^\dag_{i\sigma}\tilde{a}_{i+\hat{b}\sigma} + {\rm H.c.}\right) P, 
\end{align}
where: (i) electrons in $d_{yz}\equiv a$ ($d_{zx}\equiv b$) orbitals
can hop only along the $b$ ($a$) direction in the $(a,b)$ plane,(ii) the tilde above the operators denotes
the fact that the hopping is allowed only in the 
constrained Hilbert space, and (iii) due to the large Hund's coupling $J_H \gg t$ in the cubic vanadates \cite{Kha01} 
we project the final states resulting from the electron hopping onto the high spin states 
using the $P$ operators in Eq. (\ref{eq:ht}).
The middle terms in Eq. (\ref{eq:3}) are the superexchange terms and are somewhat lengthy \cite{Ole05},
\begin{align}\label{eq:hj123}
H^{(1)}_J&=-\frac{1}{6}Jr_1 \sum_{\langle i, j \rangle} \left({\bf S}_i \cdot {\bf S}_j + 2\right) \left(\frac{1}{4}-T^z_i T^z_j\right), \\
H^{(2)}_J&=\frac{1}{8}J \sum_{\langle i, j \rangle} \left({\bf S}_i \cdot {\bf S}_j - 1\right) \left(\frac{19}{12}
\mp \frac{1}{2}T^z_i \mp \frac{1}{2} T^z_j -\frac{1}{3} T^z_i T^z_j\right), \\
H^{(3)}_J&=\frac{1}{8}J r_3 \sum_{\langle i, j \rangle} \left({\bf S}_i \cdot {\bf S}_j-1\right) \left(\frac{5}{4}
\mp \frac{1}{2}T^z_i \mp \frac{1}{2} T^z_j + T^z_i T^z_j\right),
\end{align}
where: ${\bf S}_i$ is a spin $S=1$ operator, $T^z_i=(\tilde{n}_{ib}-\tilde{n}_{ia})/2$ is a pseudospin $T=1/2$ operator, 
and the superexchange constant $J=4t^2/U$ with $U$ being the repulsion between electrons
on the same site and in the same orbital and with $t\ll U$ being the effective hopping between
the V ions. The factors $r_1=1/(1-3\eta)$ and $r_3=1/(1+2\eta)$ (where $\eta=J_H/U$)
account for the Hund's coupling $J_H$ and originate from the energy splitting of various
$d^3$ excited states due to the various possible spin and orbital configurations \cite{Kha01}.
The last term is the three-site term which would contribute to the free hole motion \cite{Woh09},
\begin{align}\label{eq:h3s}
H_{3s}\!\! =\!\!& - \frac{1}{12} J \left(r_1+2\right)
\sum_{i,\sigma} P \left( \tilde{b}^\dag_{i-\hat{a}\sigma} \tilde{n}_{ia\bar{\sigma}} 
\tilde{b}_{i+\hat{a}\sigma}+ {\rm H. c.} \right) P \nonumber \\
&-  \frac{1}{12} J \left(r_1+2\right)
\sum_{i,\sigma} P \left( \tilde{a}^\dag_{i-\hat{b}\sigma} \tilde{n}_{ib\bar{\sigma}} 
\tilde{a}_{i+\hat{b}\sigma} + {\rm H. c.} \right) P.
\end{align}

\begin{figure}[t]
\includegraphics[width=0.46\textwidth]{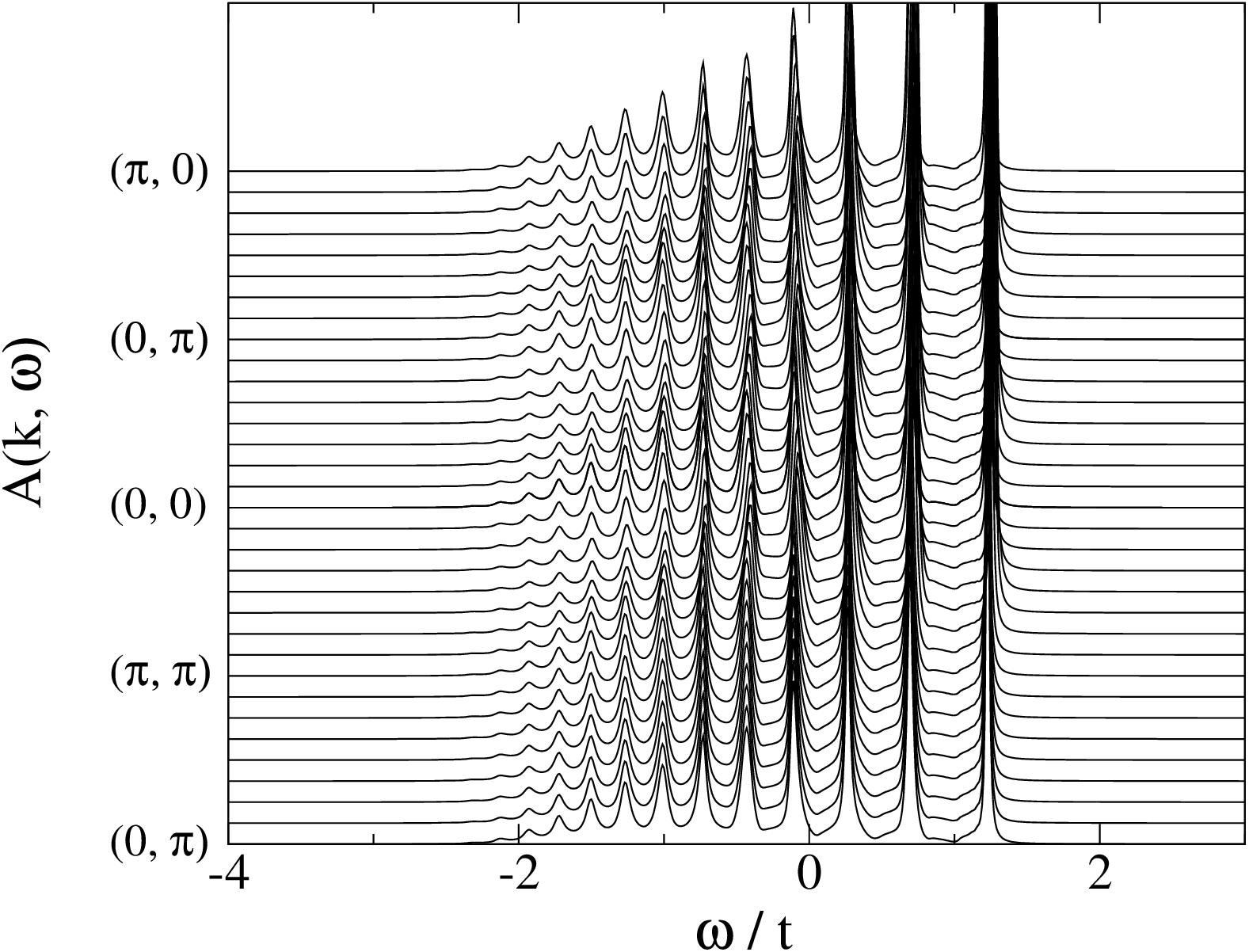}
\caption{\small{
Spectral density $A({\bf k}, \omega)$ as obtained for the $a$ orbitals 
of the model Eq. (\ref{eq:3}) with $J=0.2 t$ and $\eta=0.15t$ along 
the particular directions of the 2D Brillouin zone.}}
\label{fig:3}
\end{figure} 

In the undoped case the ground state of the model is the 2D AF and AO ordered state \cite{Kha01}.
When one hole is doped to the system the model Eq. (\ref{eq:3}) can again be expressed
in the polaron language: however, this time the hole couples {\it both} to orbiton
and magnon excitation simultaneously \cite{Woh09}. Thus, the SCBA equations are more complicated
and require an additional sum over the 2D Brillouin zone, similarly as in the case of the
coupling between a hole and two magnons \cite{Bal95}. Nevertheless, it is possible
to solve them numerically also on  a mesh of $16 \times 16$ points
-- the results for the realistic case of $J=0.2t$ and $\eta=0.15$
are shown in Fig. \ref{fig:3}.

As in the purely spin or orbital case, described in the preceding chapters, 
a well-developed quasiparticle peak on the right hand side 
of the spectrum suggests formation of the polaron also in the present case. This time
it is a {\it spin-orbital} polaron since the hole couples both to the orbitons and magnons.
Surprisingly, the quasiparticle peak has only a very small dispersion and the 
excited states reproduce almost exactly the ladder spectrum
of the purely classical spin case \cite{Mar91}. 
Since in the model Eq. (\ref{eq:3})
only the orbital (pseudo)spins are Ising-type this means that these are
the orbital degrees of freedom which are responsible for the observed classical 
behaviour.

\section{Conclusions and final discussion}

In conclusion, we studied a problem of a single hole 
doped into the half-filled ground state of the three different
cases of the $t$-$J$ model: (i) the spin model relevant for
the cuprates, (ii) the $e_g$ orbital model relevant for the
manganites, and (iii) the spin-orbital model relevant for the
vanadates. In all these three cases the hole moves by dressing 
up with the collective excitations of the ground state and forms a polaron. 

However, there are striking differences between the discussed
here polarons. On one hand, in the spin case the quasiparticle peak has
a large dispersion and the excited spectrum does not resemble the classical
ladder spectrum \cite{Mar91} at all. On the other hand, both in the orbital and in the spin-orbital
case the quasiparticle has a very tiny dispersion and the rest of the spectrum
resembles almost exactly the ladder spectrum of the classical Ising model \cite{note}.
Possibly this is one of the reasons why the
ordered state disappears very quickly with hole doping in the cuprates
whereas it is relatively stable in the manganites or vanadates:
in the two latter cases the polarons are more classical and quantum
fluctuations would not destroy the ordered state so easily.

\begin{theacknowledgments}
I would like to thank the organising committee of the course for their financial support.
I thank Andrzej M. Ole\'s, Maria Daghofer and Peter Horsch for the extremely 
fruitful discussion during the common work on this subject. I am also particularly 
grateful to Andrzej M. Ole\'s for his invaluable help and ideas. 
This work was supported in part by the Foundation for Polish Science (FNP) 
and by the Polish Ministry of Science under grant No. N202 068 32/1481.

\end{theacknowledgments}
\bibliographystyle{aipproc}   
\bibliography{sample}

\begin{thebibliography}{1}
\bibitem{note} This is because the orbital pseudospins are 
               more classical than the spins as the interaction between 
               the earlier is of the Ising-type.
\end{thebibliography}

\end{document}